\pdfoutput=1

\RequirePackage{fix-cm}

\documentclass[twocolumn]{svjour3}

\newcommand{\comm}[1]{}

\smartqed

\usepackage{graphicx,amssymb}
\usepackage{epsf,psfig}
\usepackage{txfonts}
\usepackage{natbib}

\usepackage[unicode]{hyperref}
\hypersetup{pdftitle = The title of my PDF, pdfauthor = My name, pdfsubject= The subject, pdfkeywords = keyword1 keyword2 keyword3}
\hypersetup{colorlinks = true, linkcolor = blue, anchorcolor = red, citecolor = blue, filecolor = red, pagecolor = red, urlcolor = blue}

\journalname{Astrophysics and Space Science}

\begin{document}

\title{Orbit classification in the planar circular Pluto-Charon system}

\author{Euaggelos E. Zotos}

\institute{Department of Physics, School of Science, \\
Aristotle University of Thessaloniki, \\
GR-541 24, Thessaloniki, Greece \\
Corresponding author's email: {evzotos@physics.auth.gr}}

\date{Received: 3 August 2015 / Accepted: 27 September 2015 / Published online: 5 October 2015}

\titlerunning{Orbit classification in the planar circular Pluto-Charon system}

\authorrunning{Euaggelos E. Zotos}

\maketitle

\begin{abstract}
We numerically investigate the orbital dynamics of a spacecraft, or a comet, or an asteroid in the Pluto-Charon system in a scattering region around Charon using the planar circular restricted three-body problem. The test particle can move in bounded orbits around Charon or escape through the necks around the Lagrangian points $L_1$ and $L_2$ or even collide with the surface of Charon. We explore four of the five possible Hill's regions configurations depending on the value of the Jacobi constant which is of course related with the total orbital energy. We conduct a thorough numerical analysis on the phase space mixing by classifying initial conditions of orbits and distinguishing between three types of motion: (i) bounded, (ii) escaping and (iii) collisional. In particular, we locate the different basins and we relate them with the corresponding spatial distributions of the escape and collision times. Our results reveal the high complexity of this planetary system. Furthermore, the numerical analysis shows a strong dependence of the properties of the considered basins with the total orbital energy, with a remarkable presence of fractal basin boundaries along all the regimes. Our results are compared with earlier ones regarding the Saturn-Titan planetary system.

\keywords{Restricted three body-problem; Escape dynamics}

\end{abstract}

\section{Introduction}
\label{intro}

Over the years the planar circular restricted three-body problem (PCRTBP) has played an essential role in several areas of celestial mechanics and dynamical astronomy (e.g., \citet{N04,N05,LHP15}). For instance, the modern applications to space flight missions are more numerous than the theoretical classical applications (see e.g., \citet{KLMR00,G01a,G01b,G01c,G01d,G04,SYC08}). Furthermore, today many issues in space dynamics are of paramount importance and interest. The applications of PCRTBP include the launching of artificial satellites in the solar system and they also form the basis of several planetary theories. Initially \citet{dAT14} investigated the orbital dynamics of the Earth-Moon system, while in a recent paper \citet{Z15b} (hereafter Paper I) we explored the orbital content in the Saturn-Titan system.

In this work we will follow the same computational methods in order to reveal the character of motion of a test particle (i.e., asteroid, comet or spacecraft) in the Pluto-Charon system. The aim of this work is to locate the different basins (composed either of bounded or escaping or collisional orbits) and also to relate them with the corresponding spatial distributions of the escape and collisional times of the orbits. Our numerical investigation takes place in both the configuration $(x,y)$-plane as well as in the $(x,C)$-plane.

One of the most well-observed multiple systems in the Solar System is Pluto's system. This planetary system consists of at least six celestial bodies: Pluto, its most massive satellite Charon, which was first observed in 1978 \citep{CH78}, and the four smaller satellites, that is Nix and Hydra, which was discovered in 2005 \citep{WSM05,WSM06}, and Kerberos (discovered in 2011) and Styx (discovered in 2012). Three of the outermost moons (Styx, Nix and Hydra) are found to be involved in a three-body Laplace-like mean motion resonance \citep{SH15}. At this point it should be emphasized that Charon's mass is roughly one tenth of Pluto's mass, thus making one of the most massive satellites in the solar system with respect to its primary. The system of Pluto-Charon is a unique system with the highest mass ratio in the solar system. Hubble Space Telescope (HST) observations of the barycentric wobble of Pluto give a mass ratio which is well above the critical value. Taking into account that the center of mass of this system is not inside Pluto they should be considered as a binary system.

\citet{C05} proposed a popular scenario according to that a collision between two massive Kuiper Belt objects leaded to the formation of the Pluto-Charon system. In the same paper the formation of the other two smaller satellites (Nix and Hydra) is also discussed. An in-situ formation (accretion from the remains of Charon's formation) or a capture are the two viable scenarios. Hydra which is the farthest satellite of Pluto moves in an orbit at only about 3\% of Hill's radius around Pluto thus making it one of the most compact systems \citep{SWS06}. At the time being these facts favor the accretion formation scenario however, only by the estimation of their densities we will have a more definitive answer.

The first precise measurement of the mass ratio of the Pluto-Charon system was provided by \citet{NOS93}, where they resolved Pluto and Charon using the HST. By measuring the motion of Pluto and Charon relative to a single background star over 3.2 days (one-half of Charon's orbital period), they computed a Charon/Pluto mass ratio, $\mu$, of $0.0837 \pm 0.0147$. This mass ratio corresponds to densities of 1.9 to 2.1 g $\rm cm^{-3}$ for Pluto and 1.0 to 1.3 g $\rm cm^{-3}$ for Charon, depending of course on the radii used. In the same vein, \citet{YOE94} used ground-based observations and a technique that modeled the individual centroids of Pluto and Charon to compute the mass ratio. Pluto-Charon and 10 field stars were observed for six nights (nearly one orbital period of Charon) at the University of Hawaii with the 2.2 m telescope on Mauna Kea. Using the positions of Pluto-Charon. and the field stars, they determined $\mu$ to be $0.1566 \pm0.0035$, significantly larger than the \citet{NOS93} result. This resulted in densities of 1.8 to 2.0 g $\rm cm^{-3}$ for Pluto and 1.8 to 2.3 g $\rm cm^{-3}$ for Charon.

In July 2015 the Pluto's system has been explored by in-situ spacecraft. Indeed the arrival of the New Horizons mission \citep{SS03} provided us many useful information regarding this remote planetary system. In particular, the probe provided observations of Pluto and its companions, surface imaging, and spectroscopy of the bodies' surface. Here it should be clarified that the probe did not orbit around Pluto's system as it only crossed this system before escaping from the boundaries of our solar system. The high resolution of the probe's observations (resolutions as high as 330 metres per pixel) is expected to improve the precision in our current estimates of the masses of the system's celestial bodies \citep{WGT07}. The first images proved that the color of Nix and Hydra is the same as that of Charon \citep{SMWS07}.

The paper is organized as follows: In Section \ref{mod} we present in detail the properties of the mathematical model. All the computational methods we used in order to obtain the classification of the orbits are described in Section \ref{cometh}. In the following Section, we conduct a thorough and systematic numerical investigation revealing the overall orbital structure of the planar circular Pluto-Charon planetary system. Our paper ends with Section \ref{disc} where the discussion is given.

\section{Presentation of the mathematical model}
\label{mod}

It would be very informative to briefly recall the basic properties of the planar circular restricted three–body problem (PCRTBP) \citep{S67}. The two main bodies, called primaries $P_1$ and $P_2$ move on circular orbits around their common center of gravity. The third body (also known as test particle) moves in the same plane under the gravitational field of the two primaries. It is assumed that the motion of the two primaries is not perturbed by the third body since the third body's mass is much smaller with respect to the masses of the two primaries. The non-dimensional masses of the two primaries are $1-\mu$ and $\mu$, where $\mu = m_2/(m_1 + m_2)$ is the mass ratio and $m_1 > m_2$. We choose as a synodic barycenter reference frame a rotating coordinate system where the origin is at the center of mass of the two primaries, while their centers $C_1$ and $C_2$ are located at $(-\mu, 0)$ and $(1-\mu,0)$, respectively.

The total time-independent effective potential function in the rotating frame is
\begin{equation}
\Omega(x,y) = \frac{(1 - \mu)}{r_1} + \frac{\mu}{r_2} + \frac{1}{2}\left( x^2  + y^2 \right),
\label{pot}
\end{equation}
where
\[
r_1 = \sqrt{\left(x + \mu\right)^2 + y^2},
\]
\begin{equation}
r_2 = \sqrt{\left(x + \mu - 1\right)^2 + y^2},
\label{dist}
\end{equation}
are the distances to the respective primaries.

According to \citet{url} the equatorial and polar radius of Pluto are equal (1185 km), so the oblateness of Pluto is $A_1 = 0$. Therefore in Eq. (\ref{pot}) we do not include the oblateness term which is present in Paper I for the Saturn-Titan system.

The scaled equations of motion describing the motion of the third body in the corotating frame read
\[
\ddot{x} - 2\dot{y} = \frac{\partial \Omega}{\partial x},
\]
\begin{equation}
\ddot{y} + 2\dot{x} = \frac{\partial \Omega}{\partial y}.
\label{eqmot}
\end{equation}
The dynamical system (\ref{eqmot}) admits the well known Jacobi integral
\begin{equation}
J(x,y,\dot{x},\dot{y}) = 2\Omega(x,y) - \left(\dot{x}^2 + \dot{y}^2 \right) = C,
\label{ham}
\end{equation}
where $\dot{x}$ and $\dot{y}$ are the velocities, while $C$ is the Jacobi constant which is conserved and defines a three-dimensional invariant manifold in the total four-dimensional phase space. Therefore, an orbit with a given value of its energy integral is restricted in its motion to regions in which $C \leq 2 \Omega(x,y)$, while all other regions are forbidden to the third body. The energy value $E$ is related with the Jacobi constant by $C = - 2E$.

The PCRTBP with oblateness has five equilibria known as Lagrangian points at which
\begin{equation}
\frac{\partial \Omega}{\partial x} = \frac{\partial \Omega}{\partial y} = 0.
\label{lps}
\end{equation}
Three of them, $L_1$, $L_2$, and $L_3$, are collinear points located on the $x$-axis, while the other two $L_4$ and $L_5$ are called triangular points and they are located on the vertices of an equilateral triangle. The central stationary point $L_1$ is a local minimum of the potential function $\Omega(x,y)$, while the stationary points $L_2$ and $L_3$ are saddle points. $L_1$ is between the two primaries, $L_2$ is at the right side of $P_2$, while $L_3$ is at the left side of $P_1$. The points $L_4$ and $L_5$ on the other hand, are local maxima of the potential function, enclosed by the banana-shaped isolines. The Jacobi constant values at $L_k, k = 1,..., 5$ are denoted by $C_k$. For the Pluto-Charon system we have $\mu$ = 0.099876695437731 \citep{url}, while the critical values of the Jacobi constant are: $C_1$ = 3.596603500516705, $C_2$ = 3.466490689645316, $C_3$ = 3.099456162534373, and $C_4 = C_5$ = 2.910098663205356. The projection of the four-dimensional phase space onto the configuration (or position) space $(x,y)$ is called the Hill's regions. The boundaries of these Hill's regions are called Zero Velocity Curves (ZVCs) because they are the locus in the configuration $(x,y)$ space where the kinetic energy vanishes.

\section{Computational methods and criteria}
\label{cometh}

For obtaining the orbital dynamics in the Pluto-Charon system, we need to define samples of initial conditions of orbits whose properties will be identified. For this purpose we consider dense, uniform grids of $1024 \times 1024$ initial conditions $(x_0, y_0)$ regularly distributed on the configuration $(x,y)$ plane inside the area allowed by the Jacobi constant $C$. Following a typical approach, all orbits are launched with initial conditions inside a certain region, called scattering region, which in our case is $x_{L_1} \leq x \leq x_{L_2}$ and $-0.6 \leq y \leq 0.6$. For all orbits $\dot{x_0} = 0$, while the value of $\dot{y_0}$ is always obtained from the Jacobi integral (\ref{ham}) as $\dot{y_0} = \dot{y}(x_0,y_0,\dot{x_0},C) > 0$. For the numerical integration of the orbits in every grid, we needed about between 1 minute and 22 hours of CPU time on a Pentium Dual-Core 2.2 GHz PC, depending on the escape, collisional and bounded rates of orbits in each case. For each initial condition, the maximum time of the numerical integration was set to be equal to 5000 dtu however, when a particle escaped or collided the numerical integration was effectively ended and proceeded to the next available initial condition.

In the PCRTBP system the configuration space extends to infinity thus making the identification of the type of motion of the third body for specific initial conditions a rather demanding task. Depending on the Jacobi constant the region around Charon can be connected or not to the Pluto realm through the neck around $L_1$ or to the exterior realm through the neck around $L_2$. Therefore initial conditions of orbits around Charon can be categorized into basins\footnote{The set of initial conditions of orbits which lead to a certain final state (escape, collision or bounded motion) is defined as a basin.}, namely: (i) the bounded basin containing orbits which remain in the Charon realm, (ii) the Pluto realm basin constituted by orbits which escape through $L_1$, (iii) the exterior realm basin corresponding to orbits that escape through $L_2$, (iv) the collisional basin or orbits which collide with Charon.

Now we need to define appropriate numerical criteria for distinguishing between these four types of motion. In order to consider a more realistic approach, we assume that Charon is a finite body, taking into account its mean radius approximately by 604 km (about $1.5709 \times 10^{-3}$ dimensionless length units). Therefore, if an orbit reaches the surface of Charon its numerical integration ends thus producing an orbit leaking in the configuration space. Furthermore, an escaping orbit to the Pluto realm must satisfy the conditions $x < x_{L_1} - \delta_1$, with $\delta_1 = 0.12$ and the third body inside the circle around the Pluto of radius $r_1 \leq |x_{L_3} - x_{P_1}|$. In the same vein, an escaping orbit to the exterior realm must fulfill the conditions $x > x_{L_2} + \delta_2$, with $\delta_2 = 0.10$, or $|y| > y_{L_5}$, or, if $x < x_{L_1} - \delta_1$, $r_1 > |x_{L_3} - x_{P_1}|$ (i.e., the third body is outside the circle just defined). Here we must clarify that the tolerances $\delta_1$ and $\delta_2$ were included in the escape criteria in an attempt to avoid that the unstable Lyapunov orbits are incorrectly classified as escaping orbits.

As stated earlier, in our numerical integrations the maximum time $t_f$ employed is 5000 dtu (dimensionless time units), corresponding to about 64.315 yr. Usually, the vast majority of orbits need considerable less time to escape from the system (obviously, the numerical integration is effectively ended when an orbit moves outside the system's disk and escapes). Nevertheless, we decided to use such a long integration time just to be sure that all orbits have enough time in order to escape. Remember, that there are the so called ``sticky orbits" which behave as regular ones during long periods of time. Here we should clarify, that orbits which do not escape or collide to Charon after a numerical integration of $5000$ dtu are considered as bounded regular orbits.

\begin{figure*}[!tH]
\centering
\resizebox{\hsize}{!}{\includegraphics{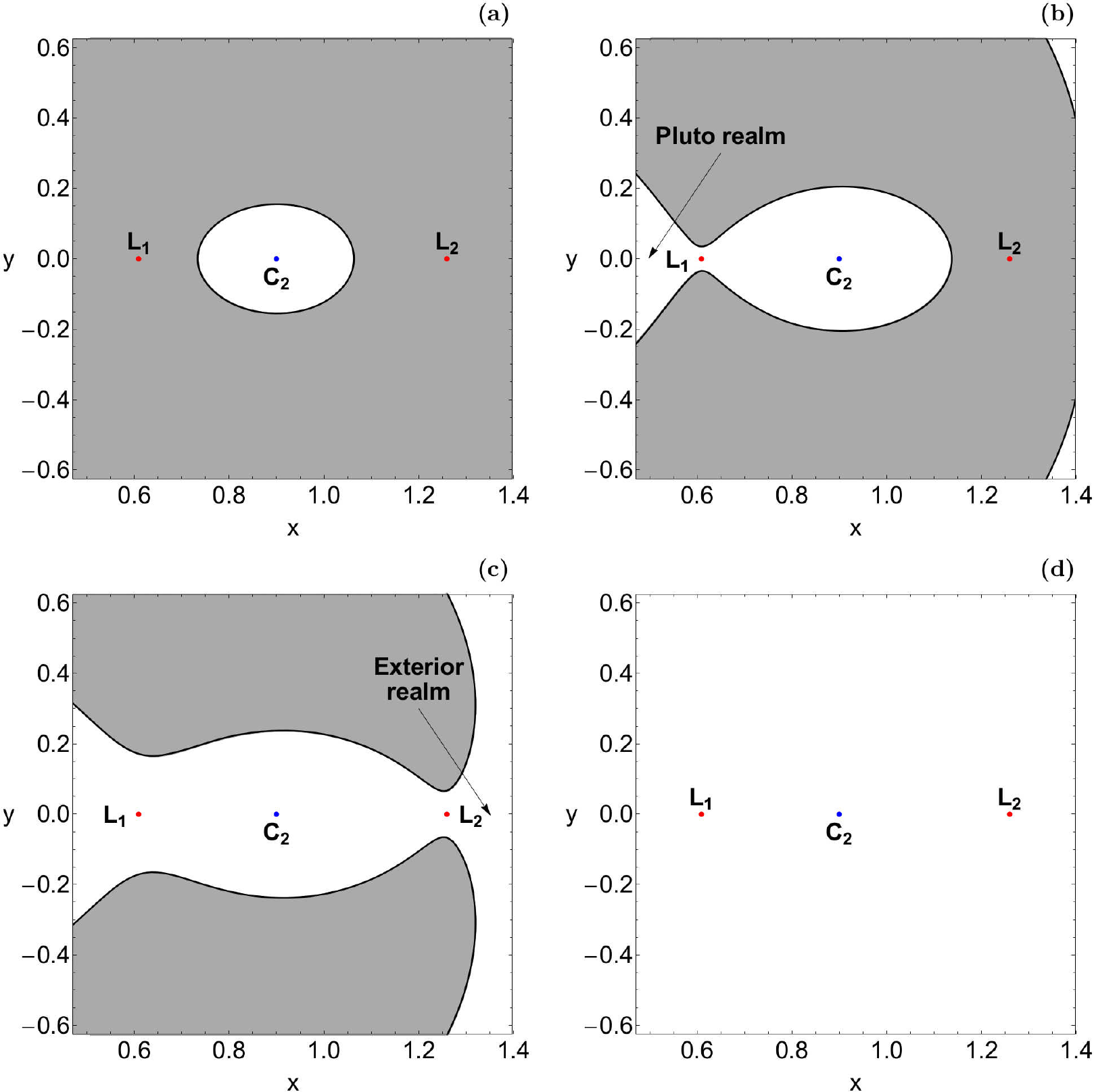}}
\caption{Four Hill's regions configurations around the vicinity of Charon. The white domains correspond to the Hill's regions, gray shaded domains indicate the forbidden regions, while the thick black lines depict the Zero Velocity Curves (ZVCs). The red dots pinpoint the position of the Lagrangian points, while the position of the center of Charon $(C_2)$ is indicated by a blue dot. (a): $C = 3.90$; (b): $C = 3.59$; (c): $C = 3.46$; (d): $C = 2.90$.}
\label{isos}
\end{figure*}

The equations of motion (\ref{eqmot}) for the initial conditions of all orbits are forwarded integrated using a double precision Bulirsch-Stoer \verb!FORTRAN 77! algorithm (e.g., \citet{PTVF92}) with a small fixed time step of order of $10^{-2}$, which is sufficient enough for the desired accuracy of our computations. Over the years, there have been numerous attempts to solve the systems of differential equations with variable time steps, however results can exhibit parametric instabilities associated with resonances between the time step variation and the orbital motion (see e.g., \citet{RF12}). Here we should emphasize, that our previous numerical experience suggests that the Bulirsch-Stoer integrator is both faster and more accurate than a double precision Runge-Kutta-Fehlberg algorithm of order 7 with Cash-Karp coefficients (e.g., \citet{DMCG12}). Throughout all our computations, the Jacobi integral (Eq. (\ref{ham})) was conserved better than one part in $10^{-11}$, although for most orbits it was better than one part in $10^{-12}$. For collisional orbits where the test body moves inside a region of radius $10^{-2}$ around Charon the Lemaitre's global regularization method is applied \citep{S67}.

\section{Numerical results - Orbit classification}
\label{numres}

The structure of the Hill's regions strongly depends on the value of the Jacobi constant. There are five distinct cases regarding the Hill's regions:
\begin{itemize}
  \item Case I: $C > C_1$: All necks are closed, so there are only bounded and collisional basins (see Fig. \ref{isos}a).
  \item Case II: $C_2 < C < C_1$: Only the neck around $L_1$ is open thus allowing orbits to enter the Pluto realm (see Fig. \ref{isos}b).
  \item Case III: $C_3 < C < C_2$: The neck around $L_2$ is open, so orbits can enter the exterior region and escape form the system (see Fig. \ref{isos}c).
  \item Case IV: $C_4 < C < C_3$: The necks around both $L_2$ and $L_3$ are open, therefore orbits are free to escape through two different escape channels.
  \item Case V: $C < C_4$: The banana-shaped forbidden regions disappear, so motion over the entire configuration $(x,y)$ space is possible (see Fig. \ref{isos}d).
\end{itemize}
In Fig. \ref{isos}(a-d) we present the structure of four possible Hill's region configurations (the fourth case where the neck around $L_3$ opens in not included because our investigation is focused only on the vicinity of Charon). We observe in Fig. \ref{isos}c the opening (exit channel) at the Lagrangian point $L_2$ through which the test particle may enter the exterior region and then leak out. In fact, we may say that these two exits (also the exit around $L_3$) act as hoses connecting the interior region of the system where $x(L_3) \leq x \leq x(L_2)$ with the ``outside world" of the exterior region.

\begin{figure*}[!tH]
\centering
\resizebox{\hsize}{!}{\includegraphics{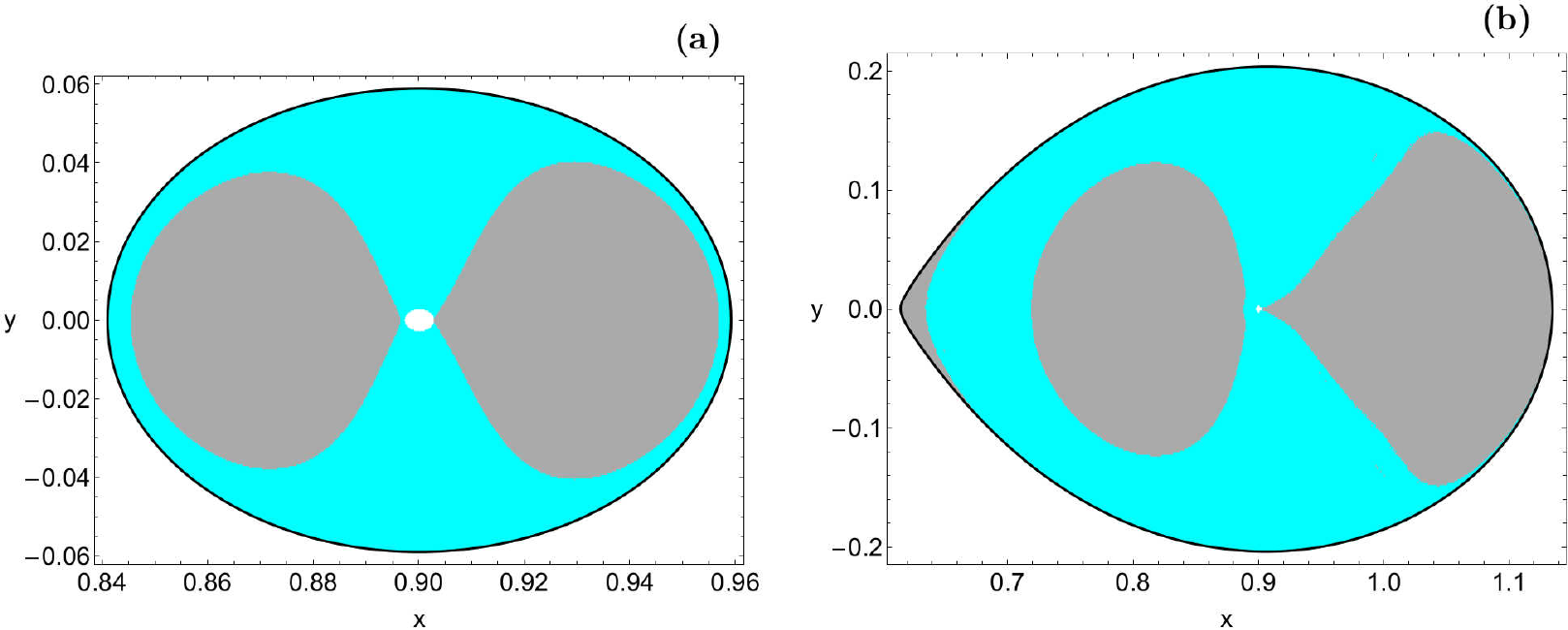}}
\caption{Basin diagrams for energy Case I. (a-left): $C = 6$, (b-right): $C = 3.597$. The color code is as follows: bounded basin of regular orbits (gray), collisional basin (cyan).}
\label{hr1}
\end{figure*}

\begin{figure*}[!tH]
\centering
\resizebox{\hsize}{!}{\includegraphics{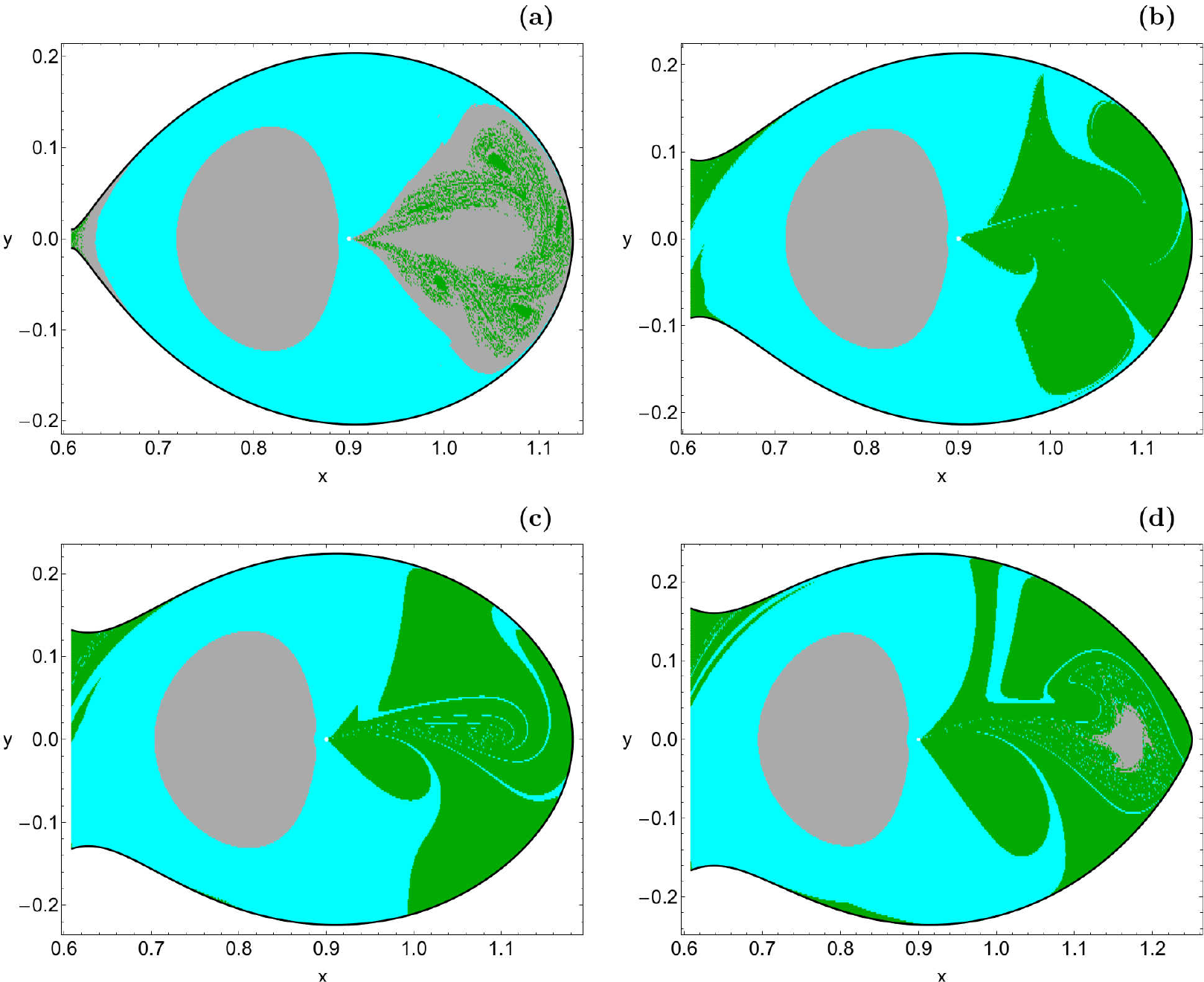}}
\caption{Basin diagrams for energy Case II. (a-upper left): $C = 3.596$, (b-upper right): $C = 3.553$, (c-lower left): $C = 3.510$ and (d-lower right): $C = 3.467$. The color code is as follows: bounded basin (gray), collisional basin (cyan), Pluto realm basin (green).}
\label{hr2}
\end{figure*}

\begin{figure*}[!tH]
\centering
\resizebox{\hsize}{!}{\includegraphics{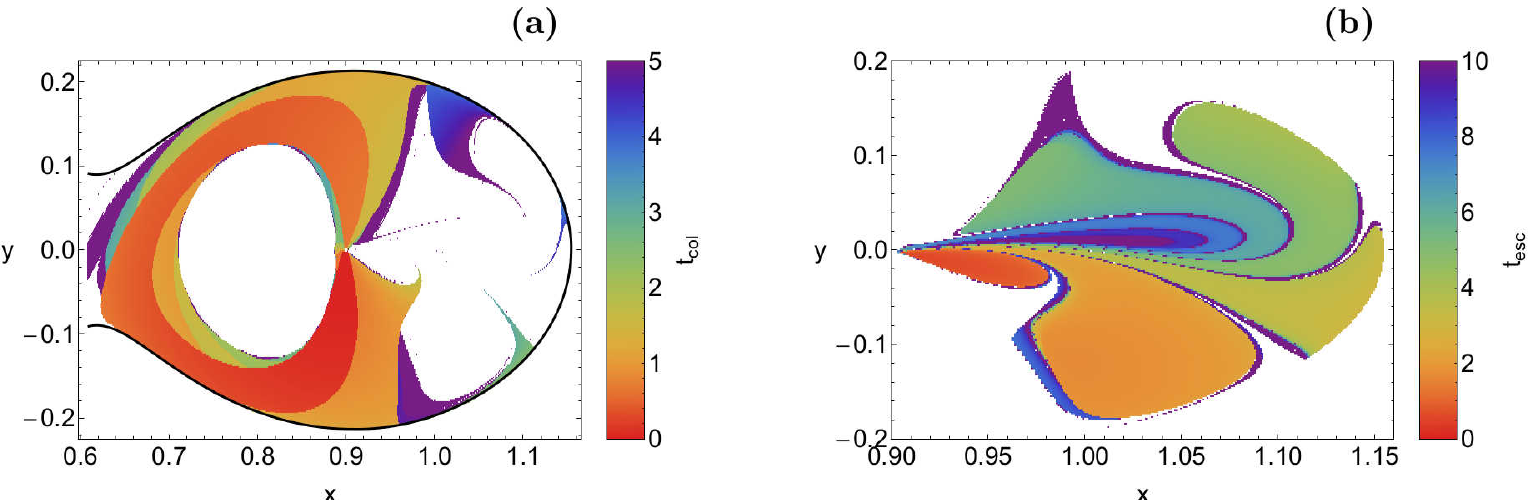}}
\caption{Distribution of the (a-left): collisional and (b-right): escape time of the orbits on the configuration $(x,y)$ space when $C = 3.553$.}
\label{tdist}
\end{figure*}

\begin{figure*}[!tH]
\centering
\resizebox{\hsize}{!}{\includegraphics{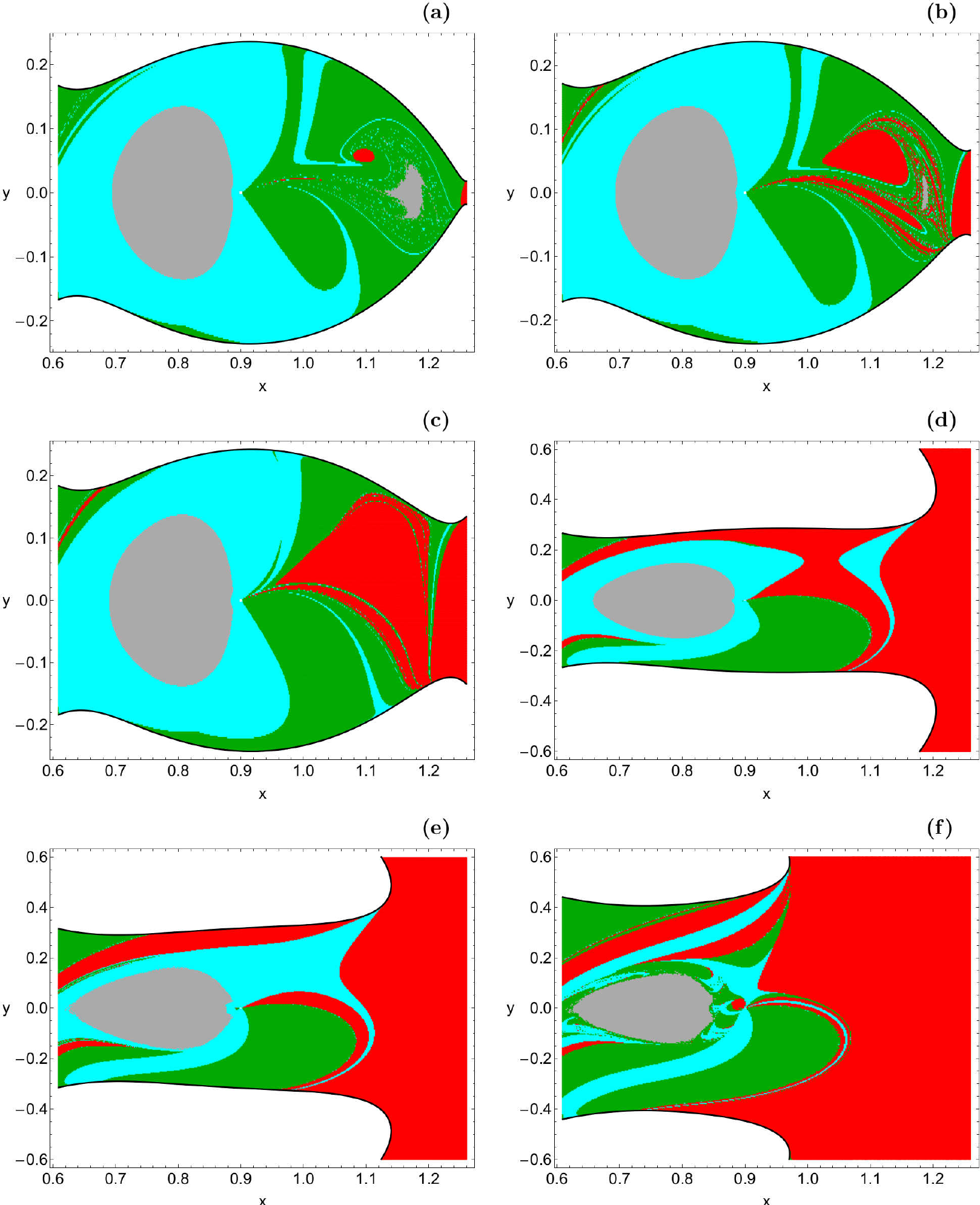}}
\caption{Basin diagrams for energy Case III. (a): $C = 3.466$, (b): $C = 3.460$, (c): $C = 3.443$, (d): $C = 3.326$, (e): $C = 3.256$, and (f): $C = 3.0995$. The color code is as follows: bounded basin (gray), collisional basin (cyan), Pluto realm basin (green), exterior realm basin (red).}
\label{hr3}
\end{figure*}

\begin{figure*}[!tH]
\centering
\resizebox{\hsize}{!}{\includegraphics{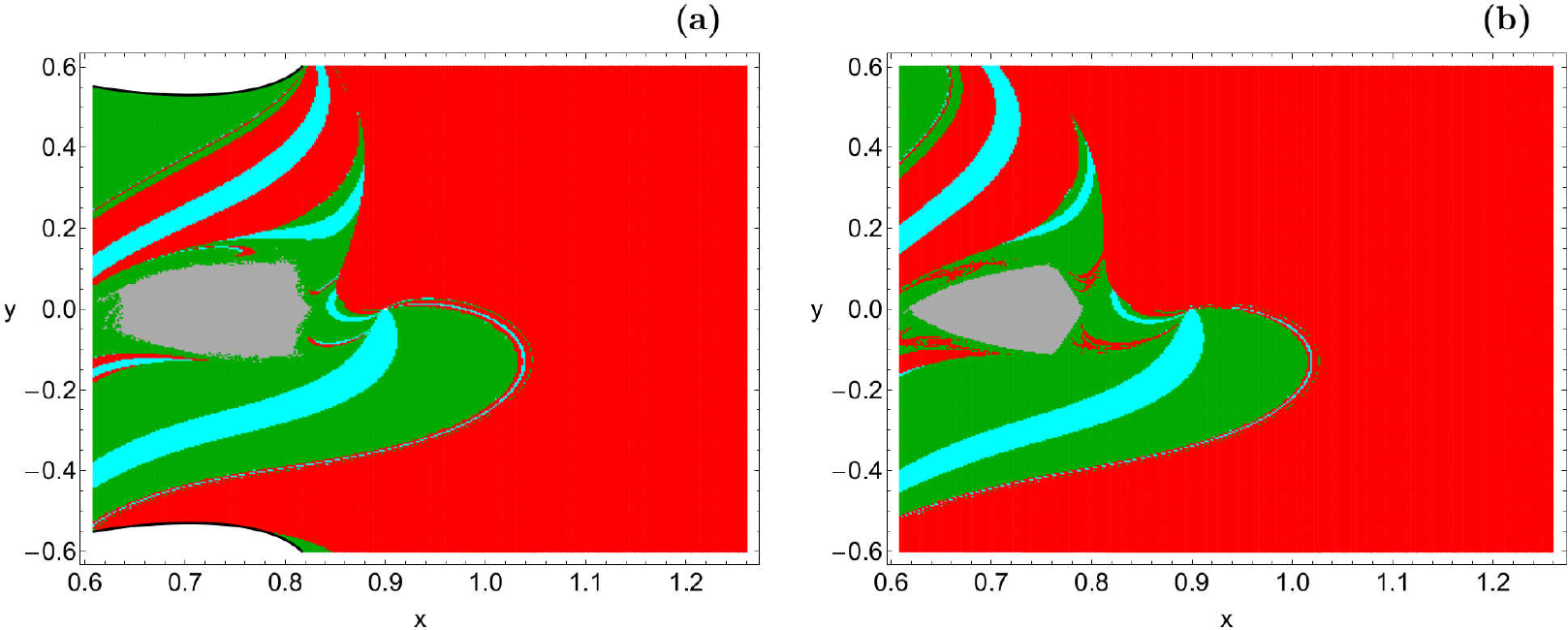}}
\caption{Basin diagrams for energy Case IV. (a-left): $C = 3$, (b-right): $C = 2.9101$. The color code is the same as in Fig. \ref{hr3}.}
\label{hr4}
\end{figure*}

\begin{figure*}[!tH]
\centering
\resizebox{\hsize}{!}{\includegraphics{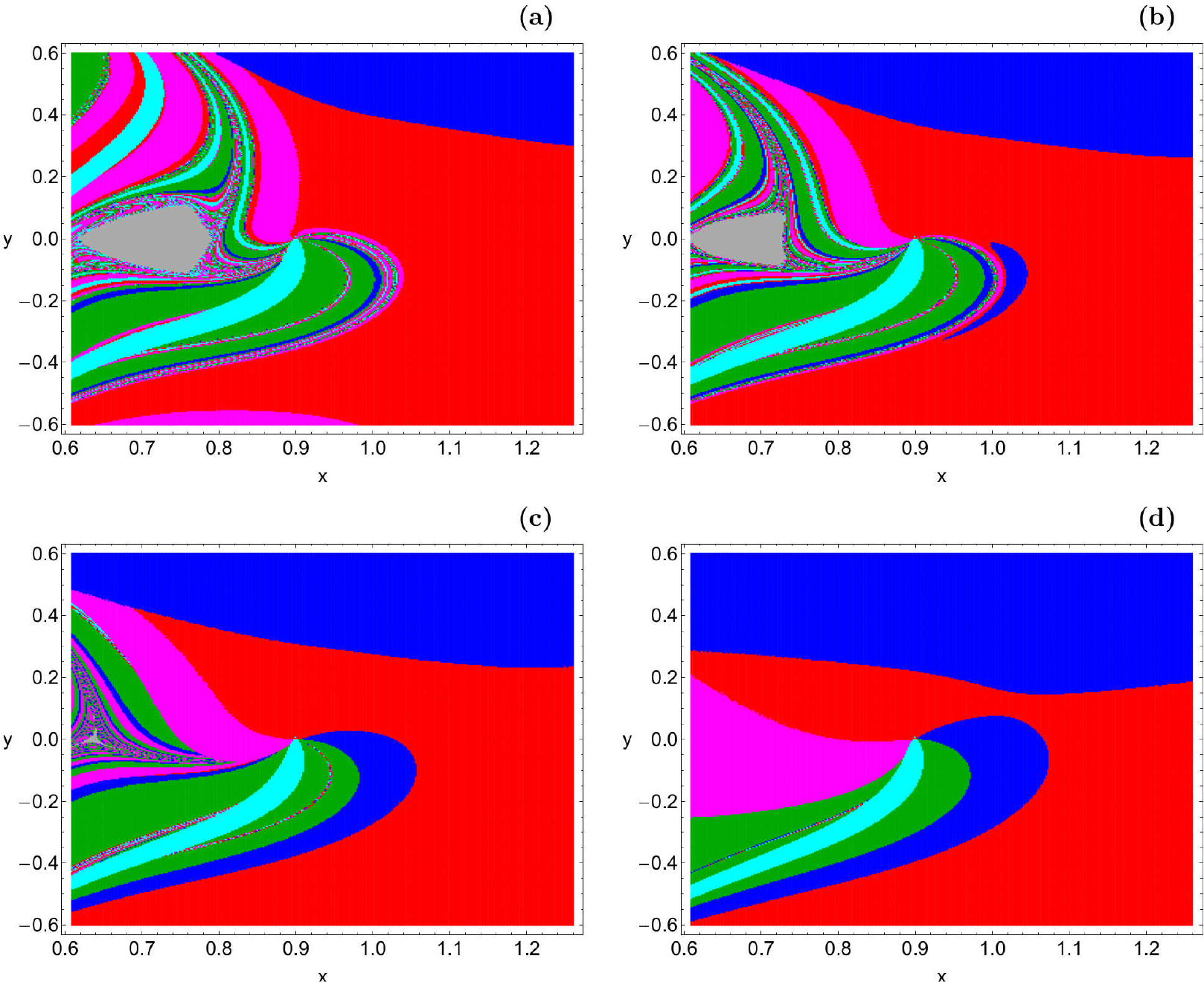}}
\caption{Basin diagrams for energy Case V. (a): $C = 2.91$, (b): $C = 2.78$, (c): $C = 2.67$, and (d): $C = 2.45$. The color code is as follows: bounded basin (gray), collisional basin (cyan), sector 1 (green), sector 2 (red), sector 3 (blue) and sector 4 (magenta).}
\label{hr5}
\end{figure*}

\subsection{Case I: $C > C_1$}

The basin diagrams for two values of the Jacobi constant are presented in Fig. \ref{hr1}(a-b). The outermost black solid line is the ZVC which is defined as $\Omega(x,y) = C/2$. Examining the two diagrams we observe that in Case I there are only bounded and collisional orbits. In particular, for $C = 6$, that Jacobi value corresponds to an energy level very close to Charon's surface, two stability islands are present inside the unified collision basin. In contrast to \citet{dAT14} and Paper I we decided to exclude initial conditions of orbits inside Charon's radius because these orbits lack of physical meaning. The white hole around the center of Charon shown in Fig. \ref{hr1}a represents these ``forbidden" initial conditions\footnote{Obviously if we numerically integrate these initial conditions we will see that they lead to immediate collision to Charon.}. The stability island situated on the right side of the Charon, surrounds a retrograde (clockwise) symmetric periodic orbit around $P_2$. On the other hand, the stability island, located on the left side of the Charon, surrounds a periodic orbit around $P_2$ which is symmetric to a reflection over the $x$-axis $(y = 0)$ and is traveled in counterclockwise sense, hence, prograde with respect to the rotating coordinate system. \citet{SS00} proved for the planar Hill problem (see also \citet{dAT14}) that the stability island on the left side of the Charon is much more stable than that on the right side in relation to $C$ variation. For $C = 3.597$ we see in Fig. \ref{hr1}b that a third smaller stability island near the left boundary of the ZVC emerges whose location is very close to Lagrangian point $L_1$. Now that the energy level corresponds to motion far away from Charon's surface the central white hole is hardly visible.

In Paper I, concerning the Saturn-Titan system, we found that for large enough values of the Jacobi constant both stability islands disappear and all the available configuration space is dominated by collisional orbits. In this system however, both stability islands survive even for extremely high values of $C$. Additional numerical calculations suggest that the behavior observed in Paper I is due to the existence of the oblateness term. Indeed in \citet{Z15a} we have shown that the oblateness coefficient has a major impact on the stability islands around both primaries. We may say that in energy case I the area of the stability islands is larger that than detected in Paper I for the Saturn-Titan system. Once more the cause of this effect is the oblateness.

\subsection{Case II: $C_1 > C > C_2$}

When $C < C_1$ the neck around $L_1$ opens thus allowing orbits to enter the Pluto realm. Indeed for $C = 3.596$ we see in Fig. \ref{hr2}a that there is an escape channel in the left side of the ZVC. Near the Lagrangian point $L_1$ we observe a tiny escape basin, while inside the right stability region we see a highly-fractal diffused structure by initial constructed by conditions of escaping orbits. For $C = 3.553$ the width of the escape channel over-doubles in size as it is seen in Fig. \ref{hr2}b. Now the right stability island has completely disappeared and a well-formed escape basin has filled it's position. As the value of $C$ decreases the width of the escape channel increases thus allowing more and more orbits to escape to the Pluto realm. In Fig. \ref{hr2}c where $C = 3.510$ one can see that inside the escape basin there are thin spiral filaments of collision orbits. This local fractality implies that a stability island is about to emerge. Indeed for $C = 3.467$ it is evident in Fig. \ref{hr2}d that a new stability island is present inside the escape basin and it is surrounded by highly fractal boundaries. As the value of the Jacobi constant decreases the size of the stability island on the left size of Charon remains almost unperturbed. In Paper I for the same energy case the escape basins were much more confined, while the size of the stability islands was smaller. Again the oblateness of Saturn is the main cause of this phenomenon.

Here it should be emphasized that the numerical integration is stopped as soon as the third body passes $L_1$ thus entering the Pluto realm. However, if we do not stop the numerical integration some orbits that initially entered the Pluto realm may return to Charon's region, or crash into on of the primary bodies, or even enter the exterior region and then escape to infinity. In our calculations we follow the approach used in \citet{dAT14} and of course in Paper I and we consider an orbit to escape to Pluto realm if the test body passes $L_1$ even if its true asymptotic behaviour at very long time limit is different.

In the following Fig. \ref{tdist}(a-b) we show how the escape and collisional times of orbits are distributed on the configuration $(x,y)$ space when $C = 3.553$. Light reddish colors correspond to fast escaping/collional orbits, dark blue/purple colors indicate large escape/collional times, while white color denote stability islands of regular motion. Inspecting the spatial distribution of various different ranges of escape time, we are able to associate medium escape time with the stable manifold of a non-attracting chaotic invariant set, which is spread out throughout this region of the chaotic sea, while the largest escape time values on the other hand, are linked with sticky motion around the stability islands. As for the collisional time we see that a small portion of orbits with initial conditions very close to the vicinity of the center of Charon collide with it almost immediately, within the first time steps of the numerical integration. In the same vein, figures depicting similar time distributions have been presented in \citet{dAT14} for the Earth-Moon system (see e.g., Figs. 7 and 10). Therefore we might conclude that the time distributions remain almost unperturbed by the change of the mass parameter $\mu$ which determines each binary system.

\subsection{Case III: $C_2 > C > C_3$}

This energy case constitutes the Hill's regions with the most interest in point of view of planetary systems. Furthermore, it has many practical applications, such as orbit determination of spacecraft mainly based on many-models and also the phenomenon of temporary capture of comets or asteroids around planets in our solar system. The exterior realm rises when $C < C_2$. In Fig. \ref{hr3}a we see that for $C = 3.466$ both necks around $L_1$ and $L_2$ are open. Once more, the majority of the configuration $(x,y)$ space is covered by initial conditions of orbits which collide to Charon. It should be noticed that a small escape basin corresponding to the exterior realm is located at the right part of the plane inside the Pluto's escape realm. When $C = 3.460$ it is seen in Fig. \ref{hr3}b that additional exterior realm basins emerge inside the Pluto's basin, while the size of the stability island in the right side of Charon is reduced and for $C = 3.443$ (Fig. \ref{hr3}c) it completely disappears. As we proceed to smaller values of the Jacobi constat (see Figs. \ref{hr3}(d-e)) the extent of the exterior realm basins grows significantly, while the collisional basin shrinks thus confined mainly near the central region of the plane. In particular we see in Fig. \ref{hr3}e that when $C = 3.256$ about 50\% of the available configuration plane is occupied by initial conditions of orbits that escape through $L_2$ to the exterior region. Finally when $C = 3.0995$ it is evident that the exterior realm basin takes over the configuration space, leading to the predominance of this type of orbits at the end of energy case III. Moreover, the extent of the collisional and the Pluto realm basins are considerably reduced, while the stability island in the left side of Charon is still present.

\subsection{Case IV: $C_3 > C > C_4$}

When $C < C_3$ the neck around $L_3$ opens thus allowing orbits to escape to the exterior region also from the left side of Pluto. Since we decided to focus our study in the vicinity of Charon the escape channel in the ZVC is not visible this time. In Fig. \ref{hr4}a we observe the orbital structure of the configuration space when $C = 3$. Things are quite similar to that discussed earlier in Fig. \ref{hr3}f. The main difference is that now the exterior realm basin covers about two thirds of the available $(x,y)$ space. As the value of the Jacobi constant decreases even further we see in Fig. \ref{hr4}b where $C = 2.9101$ that the area of the stability island on the left side of Charon has been decreased, while at the same time the growth of the exterior realm basin continues thus reducing the extent of the collisional and the Pluto realm basins.

\subsection{Case V: $C < C_4$}

In this energy case the forbidden regions completely disappear and the entire $(x,y)$ space is available for motion for the test particle. Therefore for $C < C_4$ we need to introduce new escape criteria. In particular, we divide the $(x,y)$ plane into four sectors according to a polar angle $\theta$ which starts counting from the positive part of the $x$-axis $(x > 0, y = 0)$ using the approach followed in \citet{dAT14}. Thus, considering that $\theta \in [0^{\circ},360^{\circ}]$, we have the first sector for $135^{\circ} \leq \theta < 225^{\circ}$, the second sector for $\theta \geq 315^{\circ}$ or $\theta < 45^{\circ}$, the third sector for $45^{\circ} \leq \theta < 135^{\circ}$ and the fourth sector for $255^{\circ} \leq \theta < 315^{\circ}$, respectively. We define the sectors in such a way so that sectors 1 and 2 to correspond to the previous Pluto and exterior realm, respectively. In Fig. \ref{hr5}(a-d) we present the structure of the physical $(x,y)$ plane where the initial conditions of the escaping orbits are marked with different color according to the four sectors.

In Fig. \ref{hr5}a where $C = 2.91$ we see that about 50\% of the tested initial conditions escape from sector 2. The boundaries around the stability islands are highly fractal\footnote{It should be emphasized that when we state that an area is fractal we simply mean that it has a fractal-like geometry without conducting any specific calculations as in \citet{AVS09}.}. However as the value of the Jacobi constant decreases it is evident that the size of the stability island also decreases and consequently the observed fractality. At the lowest studied Jacobi value, that is when $C = 2.45$ one may observes in Fig. \ref{hr5}d that the stability island disappears, while there is no evidence of fractal regions since all the configuration space is occupied by well-formed basins. At this point we would like to note two things: (i) at relatively low values of $C$, or in other words at large enough values of the total orbital energy, the area of sector 2 decreases, while at the same time the area of sector 3 grows. For instance at $C = 2.45$ both sectors share about 80\% of the configuration space (ii) even for small enough values of $C$ a spiral collisional basin still survives.

\subsection{An overview analysis}

\begin{figure}[!tH]
\includegraphics[width=\hsize]{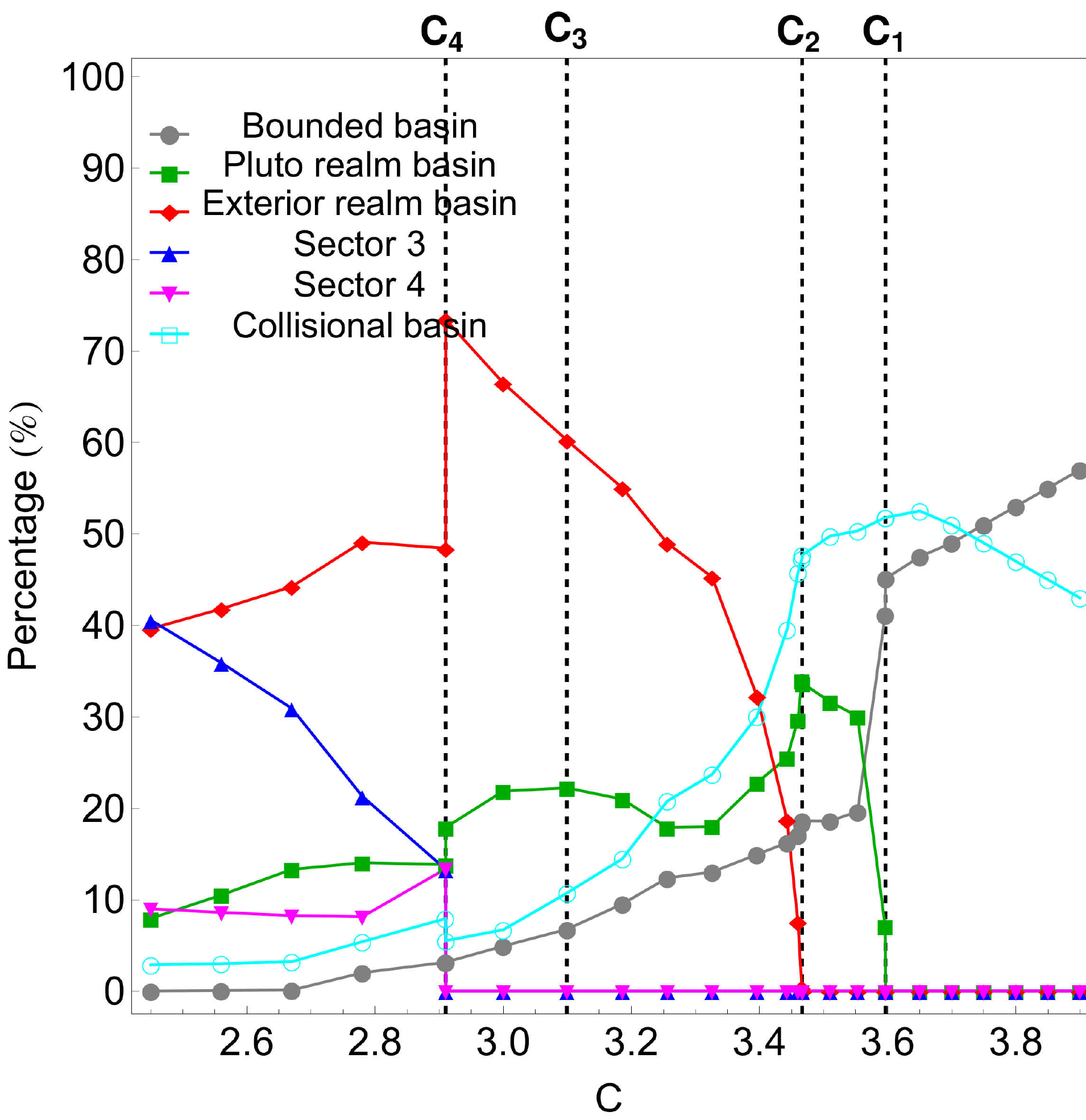}
\caption{Evolution of the percentages of the initial conditions of each considered basin as a function of the Jacobi constant. The vertical dashed black lines indicate the four critical values of $C$.}
\label{percs1}
\end{figure}

It would be very useful to monitor the evolution of the percentages of all types of orbits as a function of the Jacobi constant. Fig. \ref{percs1} shows such a diagram, where the vertical dashed black lines indicate the four critical values of the Jacobi constant $C$. We observe that at high enough values of the Jacobi constat, that is for $C > 3.90$, the regular bounded orbits is the most populated family occupying more than 60\% of the configuration space. However for $C < C_1$ the percentage of bounded orbits exhibits a sharp decrease until $C = 3.55$, while for lower values of $C$ it reduces steadily and for $C < 2.7$ it completely disappears. As the neck through $L_1$ opens for $C < C_1$ the rate of escaping orbit to the Pluto realm increases until $C = C_2$. For lower values of $C$ the trend is reversed and for low enough values of the Jacobi constant $(C < C_4)$ escaping orbits to Pluto realm possess rates around 10\%. For $C < C_2$ the second channel around $L_2$ also opens and it is seen that the percentage of orbits escaping to the exterior realm displays a rapid increase until $C = C_4$. For lower values of $C$ the rate suddenly drops and for $C < C_4$ the percentage fluctuates between 40\% and 50\% however being always the dominant type of orbits. When $C < C_4$ the percentages of sectors 3 and 4 are also present. The percentage of sector 3 displays a rather linear increase and for $C = 2.45$ its rate coincides with that of sector 2, thus sharing 80\% of the plane. On the other hand the rate of sector 4 remains almost unperturbed at around 10\%. As for the collisional basin it dominates in the interval $C_2 < C < C_1$, while for lower values of $C$ it exhibits a similar pattern to that of the bounded basin. It should be noted however, that at relatively low values of the Jacobi constant the rate of collisional orbits saturates at around 3\%, while we have strong numerical evidence than even for extremely low values of $C$ it does not become equal to zero.

\begin{figure*}[!tH]
\centering
\resizebox{\hsize}{!}{\includegraphics{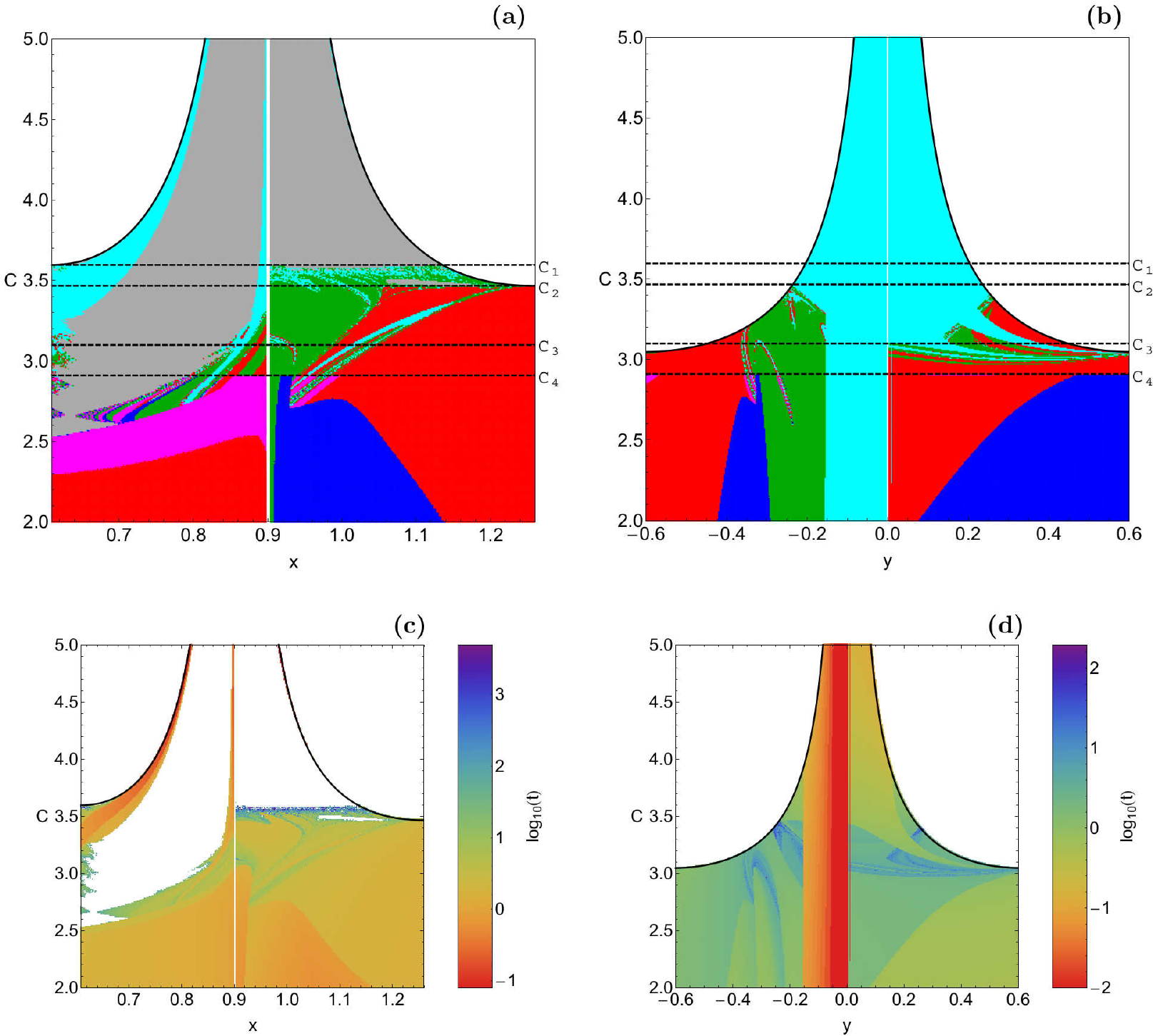}}
\caption{Orbital structure of the (a-upper left): $(x,C)$-plane; and (b-upper right): $(y,C)$-plane; (c-lower left and d-lower right): the distribution of the corresponding escape and collisional times of the orbits. The color code is the same as in Fig. \ref{hr5}.}
\label{xytc}
\end{figure*}

The color-coded diagrams in the configuration $(x,y)$ space provide sufficient information on the phase space mixing however, for only a fixed value of the Jacobi constant and also for orbits that traverse the surface of section either directly (progradely) or retrogradely. H\'{e}non back in the late 60s \citep{H69}, in order to overcome these limitations introduced a new type of plane which can provide information about areas of bounded and unbounded motion using the section $y = \dot{x} = 0$, $\dot{y} > 0$ (see also \citet{BBS08}). In other words, all the initial conditions of the orbits of the test particles are launched from the $x$-axis with $x = x_0$, parallel to the $y$-axis $(y = 0)$. Consequently, in contrast to the previously discussed type of plane, only orbits with pericenters on the $x$-axis are included and therefore, the value of the Jacobi constant $C$ can now be used as an ordinate. In this way, we can monitor how the variation on $C$ influences the overall orbital structure the Pluto-Charon system using a continuous spectrum of Jacobi constant values rather than few discrete levels. In Fig. \ref{xytc}a we present the orbital structure of the $(x,C)$ plane when $C \in [2,5]$. The four critical values of $C$ are indicated by the horizontal dashed black lines.

\begin{figure*}[!tH]
\centering
\resizebox{0.9\hsize}{!}{\includegraphics{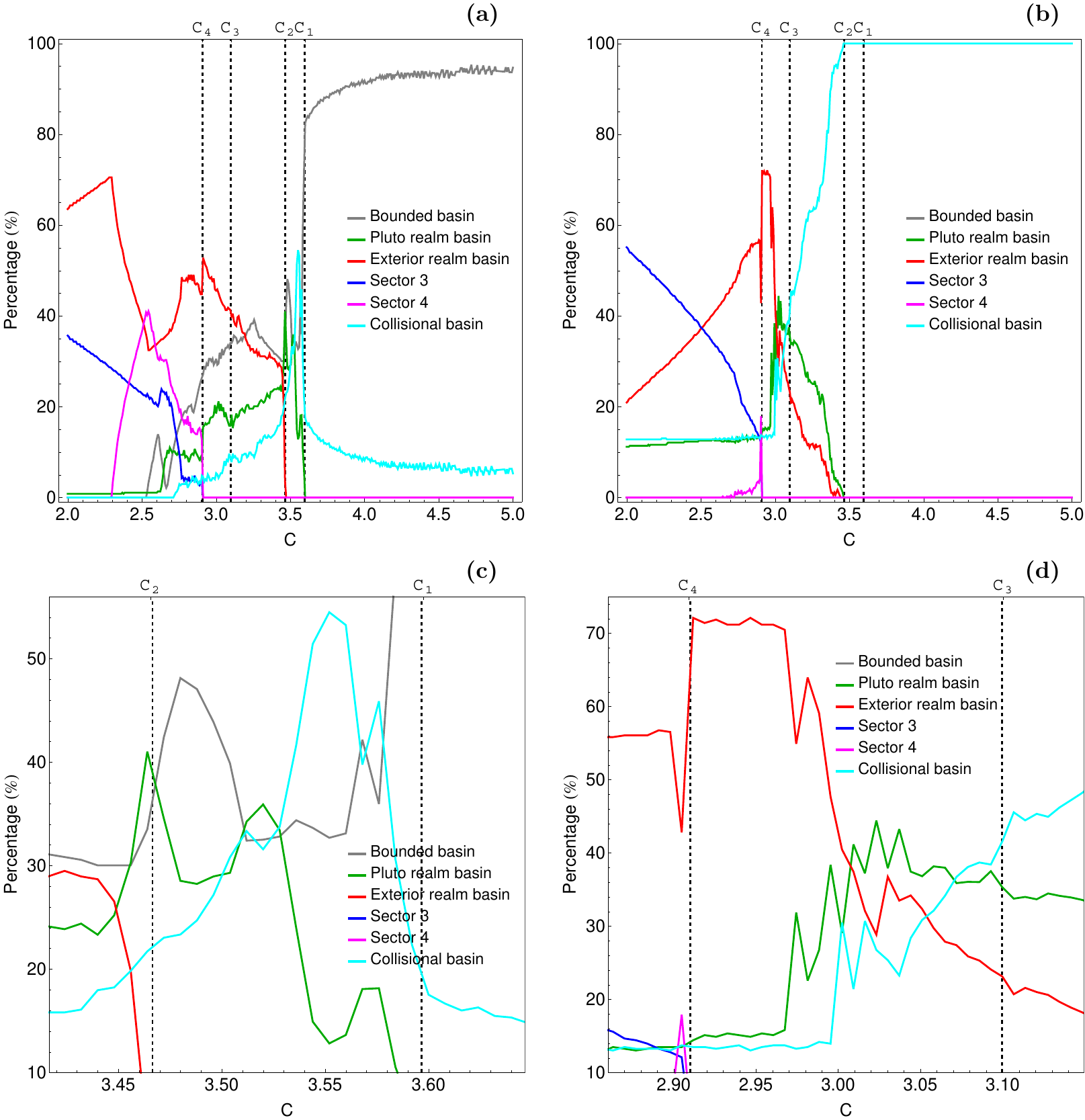}}
\caption{Evolution of the percentages of all types of orbits on the (a- upper left): $(x,C)$-plane and (b- upper right): $(y,C)$-plane as a function of the Jacobi constant. (c-lower left): zoom of panel (a) between $C_2$ and $C_1$, (d-lower right): zoom of panel (b) between $C_4$ and $C_3$. The vertical dashed black lines denote the four critical values of $C$.}
\label{percs2}
\end{figure*}

The two stability islands of prograde and retrograde motion in both sides of Charon are now visible. Here it should be emphasized that around the center of Charon there is a horizontal white lane. This lane is composed of initial conditions of orbits inside the radius of Charon and as in all previous cases these initial conditions are excluded. Below the right stability island which ends at about $C = C_1$ there is a fractal mixture of basins corresponding to escaping orbits to the Pluto and the exterior realm. Once more it is evident that for low values of the Jacobi constant the vast majority of the orbits escape to the exterior realm through $L_2$, while for high enough values of $C$ on the other hand all the initial conditions lead either to collisional or bounded motion. Furthermore, the diagram shown in Fig. \ref{xytc}a shows us how the fractality of the several basin boundaries strongly varies not only as a function of the Jacobi constant but also of the spatial variable. In particular, one can observe a very interesting phenomenon. Being more precise we see that the fractality of the basin boundaries, which is related to the unpredictability, migrates from the upper right side of Charon for relatively high $C$ (or in other words low orbital energy) to the lower left side of the smaller primary for low values of the Jacobi constant (high values of the orbital energy). It also seen that the main stability island on the left side of Charon disappears for $C < 2.5$. In general terms we may argue that the results presented in Fig. \ref{xytc}a are very similar to those of the Saturn-Titan system of Paper I. In Fig. \ref{xytc}c we illustrate how the escape and crash times of orbits are distributed on the $(x,C)$ plane. We note that the scale on the color bar is logarithmic.

In order to obtain a more complete view on how the nature of orbits in the Pluto-Charon system is influenced by the energy parameter, we follow a similar numerical approach to that explained before but now all orbits are initiated from the vertical $y$-axis with $y = y_0$. In particular, this time we use the section $x = \dot{y} = 0$, $\dot{x} > 0$, launching orbits parallel to the $x$-axis. This allow us to construct again a two-dimensional (2D) plane in which the $y$ coordinate of orbits is the abscissa, while the value of the Jacobi constant is the ordinate. The orbital structure of the $(y,C)$-plane for the same range of values of $C$ is shown in Fig. \ref{xytc}b. The black solid line is the limiting curve which distinguishes between regions of allowed and forbidden motion and is defined as
\begin{equation}
f_L(y,C) = \Omega(x = 0,y) = C/2.
\label{zvc}
\end{equation}
A very complicated orbital structure is reveled in the $(y,C)$-plane which however, in general terms, is very similar with that of the $(x,C)$-plane. One may observe that for $C < C_4$ the forbidden regions of motion around $L_4$ and $L_5$ completely disappear. With respect to the corresponding diagram of Paper I there is a major difference here. In this case there is no evidence of bounded motion in the $(y,C)$-plane and this is probably due to the absence of the oblateness coefficient. The corresponding escape and collisional times of the orbits are given in Fig. \ref{xytc}d. Looking carefully Figs. \ref{xytc}(c-d) we may conclude that the smallest escape/collisonal periods correspond to orbits with initial conditions inside the escape/collison basins, while orbits initiated in the fractal regions of the planes or near the boundaries of stability islands have the highest escape rates. Furthermore, it is seen in Fig. \ref{xytc}d that at the left side of Charon's center there is a vertical lane with initial conditions of orbits that collide to Charon within the first step of the numerical integration.

Finally the evolution of the percentages of all types of orbits on the $(x,C)$ and $(y,C)$ planes as a function of the Jacobi constant are presented in Fig. \ref{percs2}(a-b). It would be very informative to briefly discuss the evolution of the percentages and we begin with the $(x,C)$-plane shown in Fig. \ref{percs2}a. When $C > C_1$ more than 80\% of the initial conditions correspond to bounded regular orbits. As the value of the Jacobi integral increases, or the total orbital energy decreases, the rate of bounded orbits seem to saturate around 95\%. In the interval between $C_2 < C < C_1$ (see the zoom in Fig. \ref{percs2}c) we observe a drastic fall of the percentage, while in the interval $C_3 < C_2$ it fluctuates. For low values of $C$ $(C < C_4)$ the rate of regular orbits drops and for $C < 2.5$ it completely disappears. The percentage of collisional orbits displays a peak around 55\% in the interval $C_2 < C < C_1$. For $C > C_1$ it gradually decreases until $C = 4.2$ where it saturates around 7\%. For $C < C_2$ it also drops but in this case it reaches zero for $C < 2.7$. When $C < C_1$ the neck around $L_1$ opens, so the percentage of the Pluto realm basin increases until about 45\% for $C = C_2$. For lower values of the Jacobi constant the rate drops and for $C < 2.6$ it vanishes. When $C < C_2$ the percentage of the exterior realm basin increases and for $C < 2.5$ it is the most populated family although for low enough values of $C$ $(C < 2.3)$ we see a minor decrease on its rate. When the forbidden regions disappear for $C < C_4$ the percentages of sector 3 and sector 4 have non zero values. In particular the rate of sector 3 gradually increases and at the lowest value of $C$ studied, that is $C = 2$, it corresponds to about 35\%. The rate of sector 4 on the other hand, increases only until $C = 2.5$ (at about 42\%), while for lower values of the Jacobi constant its value decreases and especially for $C < 2.3$ it is effectively zero.

As it was derived form Fig. \ref{xytc}b in the $(y,C)$-plane there is no indication of bounded regular motion. In Fig. \ref{percs2}b we see that for $C > C_2$ all the integrated initial conditions of orbits lead to collision to Charon. However for $C < C_2$ the rate of collisional orbits drops significantly and for $C < C_4$ it saturates around 14\%. The percentage of the Pluto realm basin grows as soon as $C < C_1$ and for $C = 3$ it displays its maximum value at about 45\%, while for lower values of the Jacobi integral its value decreases and for $C < C_4$ it remains almost unperturbed at about 12\%. The percentage of the exterior realm basin follows a similar pattern. In particular, it grows for $C < C_2$ and at $C = C_4$ it maximizes at about 72\%, while for lower values of $C$ it drops following an almost linear trend. The rate of sector 3 grows steadily for $C < C_4$ and at the lowest studied value of the Jacobi integral is about 55\%. On the other hand, the percentage of sector 4 exhibits only a peak at about 18\% for $C = C_4$, while for all other values of $C$ its value is extremely small (lower than 5\%). Therefore, one may reasonably concludes that at low energy levels (or high values of $C$) which correspond to close ZVCs around Charon collisional motion dominates, while at high energy levels on the other hand, the vast majority of orbits escapes mainly from sector 3.

\section{Discussion}
\label{disc}

In this work we used the planar version of the circular restricted three-body problem in order to numerically investigate the orbital dynamics of a small body (spacecraft, comet or asteroid) under the influence of the potential of the Pluto-Charon planetary system. In particular, we followed the investigation of Paper I regarding the Saturn-Titan system. All the initial conditions of orbits were initiated in the vicinity of Charon which was our scattering region. We managed to determine four basins corresponding to: (i) bounded orbits around Charon, (ii) escaping orbits to Pluto realm through $L_1$, (iii) escaping orbits to the exterior realm through $L_2$, (iv) leaking orbits due to collisions with the surface of Charon. The orbital structure of the basins was monitored by varying the value of the Jacobi constant. As far as we know, this is the first time that the orbital content in the vicinity of Charon is explored in such a detailed and systematic way.

We believe that our results could be useful in future space mission design (like the New Horizons interplanetary space probe). Nevertheless, we have to stress out that in real space missions there are several types of plane perturbations mainly due to the presence of other heavy celestial bodies (e.g., planets). Near the vicinity of Charon (which was the scattering region in our work) however, all these perturbations are extremely weak and therefore negligible, so we may claim that the presented numerical results are structurally stable against out of plane perturbations.

Our numerical exploration is similar with that presented in Paper I and also in \citet{dAT14}. Therefore, it would be very informative to present the main similarities and differences between these works apart from the obvious difference of being different planetary systems. Comparing our results with that of \citet{dAT14} we conclude that, in general terms, the outcomes (both the structure and the evolution of the several basins with respect to the value of the Jacobi constant) are very similar to those reported for the Earth-Moon system. On this basis, we may argue that the particular value of the mass ratio $\mu$ of the planetary system does not practical influence the qualitative orbital content of the binary system.

On the other hand, we identified many difference between Pluto-Charon and Saturn-Titan systems. The most important of them are the following:
\begin{enumerate}
 \item In the Hill's regions configurations where $C > C_1$ we found that in the Saturn-Titan system for relatively high values of the Jacobi constant there is no indication of bounded regular motion, while the entire configuration space is covered by initial conditions corresponding to orbits that collide to Titan. In the Pluto-Charon system bounded basins is always present, while the size of the stability islands are larger in this planetary system.
 \item For $C_1 > C > C_2$, that is the second Hill's regions configurations, we identified in the Pluto-Charon system broad well-defined basins composed of initial conditions of orbits that escape to Charon's realm. These escape basins grow in size as we proceed to lower values of $C$. On the contrary for the Saturn-Titan system the presence of escape basins is very week and limited to thin filaments of initial conditions of orbits.
 \item In the third Hill's regions configurations for $C_2 > C > C_3$ the structure of the several basins between the two planetary systems is quite similar only for very low values of the Jacobi constant, or in other words for very high values of the total orbital energy. However even there we see that the size of the collisional basins is larger in the Saturn-Titan system, wile the Pluto realm basin is larger than the Saturn realm basin. Another noticeable difference is the fact the the stability island looks more prominent in the Pluto-Charon system.
 \item In the energy range $C_3 > C > C_4$ (fourth Hill's regions configurations) the regions of bounded regular motion disappear very quickly in the Saturn-Titan system, while in the Pluto-Charon system the bounded basins survive throughout the energy range. Once more the Pluto realm basins are considerably larger than the Saturn realm basins.
 \item In the last Hill's regions configurations, that is for $C < C_4$, in both systems the basin of initial conditions of orbits that escape to the exterior region dominates in the configuration space. Furthermore, another similarity is the fact that the fractality of the $(x,y)$-plane is constantly reduced with decreasing $C$. This could be explained, in a way, if we remember that the fractal regions are usually present in the vicinity of stability islands. Thus whenever the stability islands disappear the fractal regions have the same fate.
 \item In the $(x,C)$-plane for the Pluto-Charon system the stability islands survive for high enough values of the Jacobi constant. In the Saturn-Titan system on the other hand, the bounded basins disappear rather quickly as we proceed to energy levels close enough to Titan. In the $(y,C)$-plane for the Pluto-Charon system there is no indication of bounded regular motion whatsoever, while in the same plane for the Saturn-Titan system we detected two stability islands. In the Pluto-Charon system, at the left side of Pluto, a Pluto realm basin is present, while in the Saturn-Titan system a smaller Saturn realm basin is present at the right side of Titan.
 \item The distribution of the escape time of orbits on both planetary systems is very similar. This is not true however, for the distribution of the collisional time. In particular, in the Saturn-Titan system we measured the shortest collisional times. In addition we found domains of initial conditions in which the corresponding orbits collide in Titan within the first time step of the numerical integration. In the Pluto-Charon system these domains of immediate collision are very confined mainly around the center of Pluto.
\end{enumerate}
We believe that all the above-mentioned differences identified between the Pluto-Charon and the Saturn-Titan systems are due to presence of the oblateness coefficient that was included only in Paper I. Our assumption should be true since the only difference in the potential function $\Omega(x,y)$ was the inclusion or not of the oblateness coefficient.

Taking into account the detailed and systematic analysis we may conclude that our numerical task has been successfully completed. We hope that the present results to be useful in the active field of orbital dynamics in the restricted three-body problem. Since our outcomes are encouraging, it is in our future plans to expand our investigation into three dimensions and also in other planetary systems.

\section*{Acknowledgments}

I would like to express my warmest thanks to the anonymous referee for the careful reading of the manuscript and for all the apt suggestions and comments which allowed us to improve both the quality and the clarity of the paper.

\end{document}